\documentstyle[PASJadd,epsf]{PASJ95}

\begin{document}
\setcounter{page}{1}

\title{Three-Dimensional Magnetohydrodynamical Accretion Flows
into Black Holes}
\author{Mami {\sc Machida} and Ryoji {\sc Matsumoto} 
\\
{\it Department of Physics, Chiba University, 1-33 Yayoi-Cho, 
  Inage-ku, Chiba 263-8522} 
\\
{\it  E-mail(MM): machida@c.chiba-u.ac.jp} 
\\
Shin {\sc Mineshige} 
\\
{\it Department of Astronomy, Graduate School of Science, Kyoto
University, Sakyo-ku, Kyoto 606-8502}}

\abst{
Outflows and convective motions in accretion flows
have been intensively discussed recently 
in the context of advection-dominated accretion flow (ADAF)
based on two-dimensional (2D) and three-dimensional (3D)
hydrodynamical simulations.
We, however, point that without proper treatments of the
disk magnetic fields, a major source of viscosity, 
one can never derive general, firm conclusions 
concerning the occurrence of outflows and convection.
We analyzed the 3D MHD numerical simulation data of 
magnetized accretion flows initially threaded by weak
toroidal magnetic fields, finding large-scale convective motions
dominating near the black hole. 
In contrast, outflows occur only temporarily and are not very
significant in our simulations.
If there grow strong vertical fields somehow, however,
formation of bi-polar jets is inevitable.
It is claimed that radiation could be dominant 
at the outermost zones of the convective disks
because of outward energy flow by convection, however,
this is no longer the case in convective MHD flows, since
accretion energy can be released via magnetic reconnection
in the inner parts.
Such reconnection leads to sporadic flare events, 
thus producing substantial variability in out-going radiation.
}
\kword{Accretion, accretion disks --- Black holes --- 
Convection --- Outflow --- Magnetohydrodynamics} 
\maketitle

\section{Introduction}

Rapid progress has been made recently in the research field
of black-hole accretion flow and
our knowledge about flow properties has been remarkably enriched.  
In addition to the standard disk (Shakura, Sunyaev 1973),
the existence of a distinct type of flow,
advection-dominated accretion flow (ADAF) 
has been widely recognized.  
Although ADAF was proposed long time ago by Ichimaru (1977),
its significance is only recently fully appreciated
(Narayan and Yi 1994, 1995; Abramowicz et al. 1995;
for a review, see Kato et al. 1998). 

Narayan and Yi (1994) made two important pieces of predictions:
possible occurrence of
convection (in the radial direction) and outflows.
Since radiative cooling is inefficient,
entropy of accreting gas should monotonically
increase towards a black hole (i.e., in the direction
of gravity), a condition for a convective instability.
Convection was later confirmed numerically
(e.g. Igumenshchev et al. 1996; Stone et al. 1999).
The occurrence of outflow is, on the other hand,
still open to question; the self-similar model points
an advection-dominated flow having positive Bernoulli parameter, $Be$, 
meaning that matter is gravitationally unbound and can form outflows
(see also Blandford, Begelman 1999).  However, 
matter with positive $Be$ may not necessarily produce
powerful unbound outflows (Abramowicz et al. 2000).
Certainly, the traditional, radially one-dimensional treatments
are inappropriate to discuss this issue.

Through extensive 2D hydrodynamical simulations,
Igumenshchev and Abramowicz (2000, hereafter IA00) have found
different modes of accretion flow realizing
in advection-dominated regimes,
depending on the magnitude of the viscosity parameter, $\alpha$.
They found 
outflows for a moderately large $\alpha$ ($\sim 1$) and convection 
and/or large-scale circulations for small $\alpha \lsim 0.1$ 
(see also Igumenshchev et al. 2000 for 3D calculations).  

\begin{figure*}[t]
\epsfxsize=19.0cm
\centerline{\epsfbox{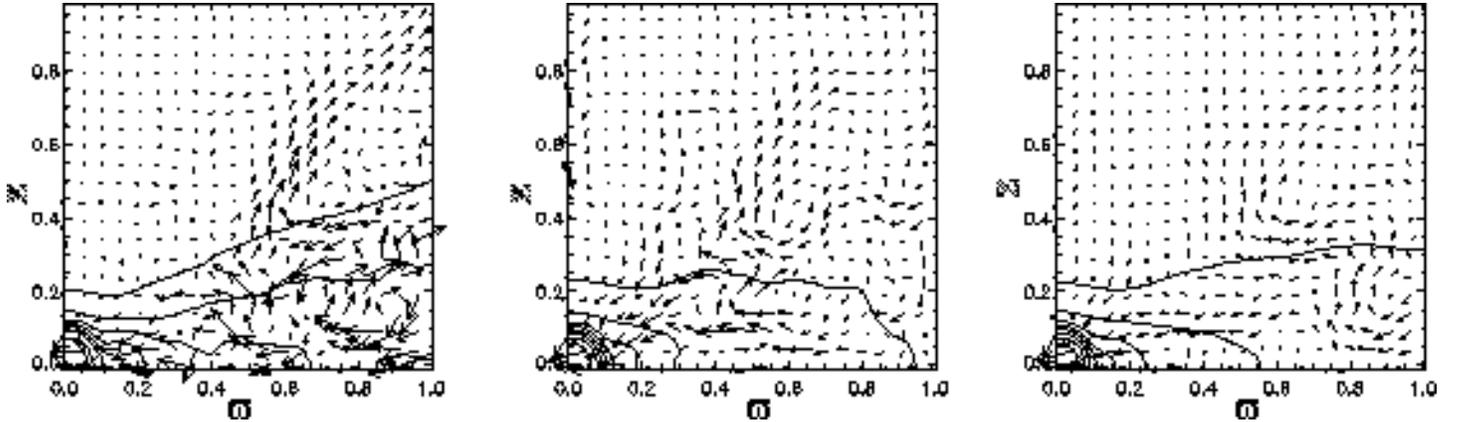}}
\caption
{Density isocontours overlaid with 
a momentum vector multiplied by $r$ (i.e. $\rho r \mbox{\boldmath $v$}$) 
 on the ($\varpi,z$) plane 
averaged over azimuthal angle ($\phi$):
(a) snapshot at $t=10.5$, (b) that at $t=19.7$, and 
(c) the time averages 
from $t=10.5$ to $t=20.8$, respectively.
The spacing of density isocontour is $\triangle\log\rho = 0.1$.
The density contours where $\rho < 0.1$ are not shown.
}
\end{figure*}

However, the
magnitude of viscosity is not a free parameter in real situations
but should be determined in a self-consistent way.
It is now widely believed that
magnetic fields provide a major source of disk viscosity. 
If so, without proper treatment of disk magnetic fields 
one can never derive solid conclusions concerning flow patterns.
Dynamics of magnetically dominated accretion is in question.
This prompted us to examine the three-dimensional data of the
global, MHD disk calculations first made by Machida et al. (2000), 
aiming to establishing a general, new view of accretion flow
in advection-dominated regimes.
The results of MHD flow structures are presented in section 2.
The final section is devoted to discussion.

\section{Three-Dimensional Flow Structure}

Machida et al. (2000)
 calculated how magnetic field evolves in a rotating 
MHD torus with its center at $\varpi=r_0$ 
initially threaded by toroidal ($B_\phi$) fields
under the Newtonian potential (Okada et al. 1989) with cylindrical
coordinate ($\varpi, \phi, z$). 
They solved ideal MHD equations, but entropy can increase
by shock heating. 
Since no cooling is taken into account in the computations, 
the simulated disk is of an ADAF type.
They assumed the adiabatic index, $\gamma = 5/3$,
and no heat conduction.
The numbers of mesh points are
$(n_{\varpi} \times n_\phi\times n_z)=(200\times 64\times 240)$.
The outer boundaries at $\varpi=6.4~r_0$ and at $z=11.8~r_0$
are free boundaries at which waves can transmit.
A periodic boundary condition is imposed for $\phi$-direction.
The grid size is $\triangle \varpi = \triangle z = 0.01~r_0$
for $0\leq \varpi \leq 1.2~r_0$ and $0\leq z\leq r_0$
and otherwise increases with $\varpi$ and $z$. 
We imposed absorbing boundary condition at $r =0.1~r_0$ where 
$r \equiv (\varpi^2 + z^2)^{1/2}$. 
To initiate non-axisymmetric evolution,
small-amplitude, random perturbations are imposed at
$t=0$ for azimuthal velocity.
Although the initial torus has a flat angular-momentum distribution,
the angular-momentum profile soon approaches that of the Keplerian
($\sim \varpi^{1/2}$) after several rotations.  
Since it is difficult to add gas threaded by magnetic fields,
we assume no mass input.
However, the flow can be regarded as being quasi-stationary 
inside $r_0$, since the total calculation time is $t \sim 22$ 
(in a unit of the rotation period at $r_0$)
which by far exceeds the mass-flow timescale ($\sim r/|v_r|$) 
in the inner zone.

As time goes on,
each component of magnetic fields is rapidly
amplified via a number of MHD instabilities together with the
differential rotation.
The maximum field strength is determined 
either by field dissipation by reconnection 
or field escape from accretion flow via Parker instability.  
As a result, the plasma $\beta$ ($\equiv p_{\rm gas}/p_{\rm mag}$, 
the ratio of gas pressure to magnetic pressure)
finally reaches $\sim 2-10$ on average, 
irrespective of the initial $\beta$.  The corresponding
$\alpha$ value is $\alpha \sim 0.01-0.1$.  
Even lower-$\beta$ ($<1$) values are observed in
the regions with filamentary shapes.   
Although the spatial distribution of the fields is
quite inhomogeneous (Kawaguchi et al. 2000), matter
distribution is somewhat smoother.
In the following, we analyze the case with 
initial $\beta_0 = 100$ at $(\varpi,z) = (r_0,0)$.  

Figure 1 displays the snapshots (left and middle) and the time averages 
(right) of density contours and momentum vectors on the $(\varpi,z)$ plane.
Note that each quantity is averaged over azimuthal angles.
Large-scale convective motions, which look very similar to those
of IA00 (their figure 15) are evident in these panels; that is,
the accretion is suppressed in the equatorial plane and 
the mass inflows concentrate mainly along the upper and lower surfaces 
of the torus-like accretion disk.
The polar inflows are highly supersonic, although
accompanying mass flux is small because of low density.
Note that we assume equatorial symmetry, thus 
large-scale motion penetrating the equatorial plane is artificially 
suppressed in our simulations, while it is allowed in IA00.

\begin{figure*}[t]
\epsfxsize=10.0cm
\centerline{\epsfbox{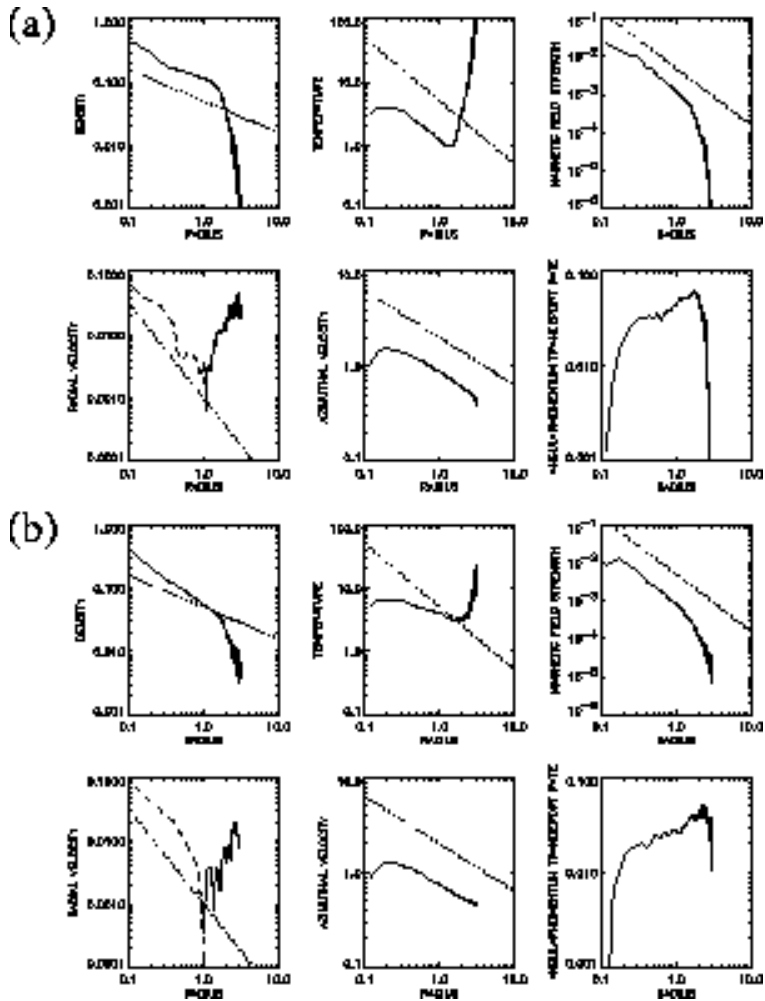}}
\caption
{
Various physical quantities averaged over angles
as functions of radius ($r$):
matter density (upper-left), 
temperature (upper-middle),
magnetic energy (upper-right),
radial velocity $v_r$ (lower-left), azimuthal velocity (lower-middle), 
and viscosity parameter, 
$\alpha \equiv -\langle B_r B_{\phi}/4\pi\rangle/\langle 
p_{\rm gas}\rangle$ 
(lower-right), respectively, at different epochs,
(a) at $t=10.5$ and (b) at $t=19,7$, respectively.
The dashed curves represent power-law relations, 
such as $\rho \propto r^{-0.5}$, $T \propto r^{-1.0}$, 
$B^2 \propto r^{-1.5}$, $v_r \propto r^{-1.5}$, and
$v_{\phi} \propto r^{-0.5}$.  
Long dashed curves in the lower-left panels indicate portions 
of negative $v_r$ (inflow).
The unit of velocity is the Keplerian rotation speed at $r_0$.}
\end{figure*}

In figure 2, we plot various physical quantities averaged over 
a spherical shell with a constant $r$:
%
\begin{equation}
\bar{q}(r) \equiv  
       \frac{\int\!\int\! q(r,\theta,\phi)\ \sin{\theta}\ d\theta\ d\phi}
          {\int\!\int\! \sin{\theta}\ d\theta\ d\phi}
\end{equation}
for $q = \rho, T, B^2, \alpha$, and 
%
\begin{equation}
\bar{\mbox{\boldmath $v$}}(r) = 
      \frac{\int\!\int\!\rho\mbox{\boldmath $v$}(r,\theta,\phi)\ 
                        \sin{\theta}\ d\theta\ d\phi}
            {\int\!\int\!\rho\sin{\theta}\ d\theta\ d\phi}.
\end{equation}
The results are summarized in Table 1, in which we also
compare radial dependences of various quantities 
for various types of accretion.  
Note that outer parts of the simulated disks 
are formed by out-going motions of the disk material 
as a result of re-distribution of the initial disk
angular momenta as is indicated by positive radial velocities. 
In the MHD flow, we concentrate on the flow structure at $r \leq r_0$,
where inflow motions dominate (indicated by long-dashed lines in
the $v_r$ distribution).

\begin{table}[b]
\small
\begin{center}
Table~1.\hspace{4pt}Various types of accretion flow.\\
\end{center}
\vspace{6pt}
\begin{tabular*}{\columnwidth}{@{\hspace{\tabcolsep}
\extracolsep{\fill}}lcccc} 
\hline\hline\\[-6pt]
 accretion mode & $\rho(r)$  & $T(r)$         & $v_r (r)$ & $B^2(r)$  \\
\hline\\[-6pt]
  ADAF     & $\propto r^{-3/2}$    & $\propto r^{-1}$ 
           & $\propto r^{-1/2}$    &   $\cdots$     \\
 outflow   & $\propto r^{-1}$      & $\propto r^{-1}$ 
           & $\propto r^{-1}$      &   $\cdots$     \\
convection & $\propto r^{-1/2}$    & $\propto r^{-1}$ 
           & $\propto r^{-3/2}$    &   $\cdots$     \\
MHD flow   & $\propto r^{-0.5}$    & $\propto r^{-1.0}$ 
           & $\propto r^{-1.5}$    & $\propto r^{-1.5}$\\
(at $t=10.5$) & & & \\ 
MHD flow   & $\propto r^{-0.8}$    & $\propto r^{-0.5}$ 
           & $\propto r^{-1.3}$    & $\propto r^{-1.6}$\\
(at $t=19.7$) & & & \\ 
\hline
\end{tabular*}
\end{table}

In general, our numerical results remarkably agree well with those of the
convective disk at least at $t=10.5$,
confirming that convection dominates the flow dynamics. 
It is interesting to note that density profiles differ at
$t=10.5$ (named as early stage) and at $t=19.7$ (late stage).
We notice that
entropy [$\propto \ln(p_{\rm gas}/\rho^\gamma)$ with $\gamma = 5/3$]
increases with decreasing $r$ in the early stage
(namely, the flow is certainly convectively unstable), whereas
it is nearly flat (i.e., the flow is isentropic) in the later stage.
This indicates that convection efficiently works in the early stage 
to smooth out entropy profile, leading to a flat entropy profile
in the late stage.
The plasma $\beta$ is roughly constant radially,
whereas electron drift velocity
($\propto j/\rho\propto |{\bf\nabla}\times {\bf B}|/\rho$),
a key factor to turn on anomalous resistivity,
increases with a decreasing $r$ as $\propto r^{-1}$.
Magnetic reconnection, hence, preferably occurs at small radii.
We also notice that $\alpha$ has a weak radial dependence; 
roughly, $\alpha \propto r^{0.5}$.

\section{Discussion}

Magnetic fields play many crucial roles in activating
accretion disks:
source of viscosity (Shakura, Sunyaev 1973),
producing rapid fluctuations (e.g. Kawaguchi et al. 2000),
jet formation (e.g. Matsumoto 1999), and so on.
Nevertheless, our knowledge about the disk magnetic fields
still remains poor mainly because of their highly nonlinear and
global nature, requiring sophisticated multi-dimensional simulations
which are only recently available.
A number of questions have been addressed: 
is the field strength strong enough
to account for observed rapid accretion?
Does convection really occur in a magnetized disk?
Or is it more likely for an ADAF to produce outflows/jets?

Although we have not run many models,
we obtain a sort of answers for some cases.
The field strength can indeed grow to be strong enough
explain observed accretion phenomena.
Convective motions are evident in the simulations, indicating
that convection may be rather general phenomena in accretion flows.
The estimated viscosity parameter is, $\alpha \sim 0.01 - 0.1$,
for which we indeed expect the onset of convection from
2D/3D hydrodynamical simulations.  
More precisely, we find $\alpha \propto r^{0.5}$.
However, we should keep in mind that
the presence of patchy low-$\beta$ regions implies
that simple $\alpha$-type prescription may not perfectly work.

As Machida et al. (2000) already noted, 
outflow is temporarily observed, although it is not very significant;
the mass outflow rate is less than the mass accretion rate 
by some factor.
Although we began calculations with weak toroidal fields,
if we start with strong vertical fields, 
magnetically-driven bi-polar jets are evident (e.g. Matsumoto
 1999). The flow pattern of MHD disks may depend on the 
initial field configurations.  
On the other hand, injection of a magnetized, differentially rotating
buoyant element naturally can drive outflows (Turner et al. 1999).
We need further MHD studies
before finally settling down the issue of convection/outflows.

It is suggested (Igumenshchev 2000) that 
since energy can be carried outward by convective motions,
radiation could be dominant at the outermost zones.  We,
however, point out that magnetic reconnection is likely to occur 
in the inner zones and to release substantial
fraction of the total energy of accretion flows in the inner zone.  
This is a very likely possibility, if indeed
magnetic flares are responsible for variability
(e.g. Kawaguchi et al. 2000).  In fact, we find large
field energy in the inner zone, which can be released in 
a sufficiently short time before
fields are brought outward by convective gas motions.
That is, the computed spectra on the basis of the ADAF model
will not need significant changes even in the presence of
large-scale circulations.

We have not yet included conductive energy transport, although
two-dimensional hydrodynamical simulations found significant
changes in flow structure caused by conduction;
any conductive models did not show bipolar outflows nor convection,
since the thermal conduction mainly acts as a
cooling agent (IA00).
It is tempting to check if this is the case in MHD disks, as well.

\par
\vspace{1pc} \par
This work was partially supported by 
Japan Science and Technology Corporation (ACT-JST) and
by the Grants-in Aid of the Ministry of Education, Science, Sports, 
and Culture of Japan (10640228, SM).
Numerical computations were carried out by using
Fujitsu VPP300/16R at National Astronomical Observatory, Japan.

\section*{References}

\re Abramowicz M. A., Chen X. M., Kato S., Lasota J.-P., 
    Regev O. 1995, ApJL, 438, 37
\re Abramowicz M.A., Lasota J.-P., Igumenshchev I.V. 2000,
    MNRAS, 314, 775
\re Blandford R.D., Begelman M.C. 1999, MNRAS 303, L1
\re Ichimaru S. 1997, ApJ, 214, 840
\re Igumenshchev I.V. 2000, MNRAS in press
\re Igumenshchev I.V., Abramowicz M.A. 2000, astro-ph/0003397 (IA00)
\re Igumenshchev I.V., Chen X., Abramowicz M.A. 1996, MNRAS 278, 236
\re Igumenshchev I.V., Abramowicz M.A., Narayan R. 2000, ApJL, 537, 27
\re Kato S., Fukue J., \& Mineshige S.1998, 
    Black-Hole Accretion Disks  (Kyoto Univ. Press, Kyoto)
\re Kawaguchi T., Mineshige S., Machida M., Matsumoto R.,
    and Shibata K. 2000, PASJ 52, L1
\re Machida M., Hayashi M. R. \& Matsumoto R.\ 2000, ApJL, 532, 67
\re Matsumoto R. 1999, Numerical Astrophysics, ed. Miyama S.,
    Tomisaka K., and Hanawa T. (Kluweer Academic Publishers), p.195
\re Narayan R., Yi I. 1994, ApJ 428, L13
\re Narayan R., Yi I. 1995, ApJ 452, 710
\re Okada R., Fukue J., Matsumoto R. 1989, PASJ 41, 133
\re Shakura N. I., Sunyaev R. A. 1973, A\&A, 24, 337
\re Stone J.M., Pringle J.E., Begelman M.C. 1999, MNRAS 310, 1002
\re Turner N. J., Bodenheimer P., Rozyczka M. 1999, ApJ, 524, 129
\end{document}